\documentclass[12pt, draftclsnofoot, onecolumn]{IEEEtran}

\usepackage{amsmath}
\usepackage{mathrsfs}
\usepackage{pifont}
\usepackage{bbding}
\usepackage{amsmath,epsfig}
\usepackage{stmaryrd}
\usepackage{amssymb}
\usepackage{amsfonts}
\usepackage{epic}
\usepackage{graphicx}
\usepackage{curves}
\usepackage{cite}

\usepackage{algorithm}

\usepackage{algpseudocode}
\usepackage{cases}
\usepackage{stfloats}
\usepackage{latexsym}
\usepackage{epstopdf}
\usepackage{epic}
\usepackage{bm}
\usepackage{multirow}
\usepackage{xcolor}
\usepackage{mathrsfs}
\usepackage{pifont}
\usepackage{bbding}
\usepackage{amsmath,epsfig}
\usepackage{mathbbold}
\usepackage{stmaryrd}
\usepackage{amssymb}
\usepackage{amsfonts}
\usepackage{epic}
\usepackage{graphicx}
\usepackage{curves}
\usepackage{algorithm}
\usepackage{algpseudocode}

\usepackage{stfloats}
\usepackage{latexsym}
\usepackage{epstopdf}
\usepackage{epic}
\usepackage{multirow}

\usepackage{subfigure}
\usepackage{color}

\usepackage{amsmath}
\usepackage{mathrsfs}
\usepackage{pifont}
\usepackage{bbding}
\usepackage{amsmath,epsfig}
\usepackage{mathbbold}
\usepackage{stmaryrd}
\usepackage{amssymb}
\usepackage{amsfonts}
\usepackage{epic}
\usepackage{graphicx}
\usepackage{curves}
\usepackage{cite}

\usepackage{algpseudocode}

\usepackage{stfloats}
\usepackage{latexsym}
\usepackage{epstopdf}
\usepackage{epic}
\usepackage{bm}
\usepackage{multirow}
\usepackage{xcolor}
\usepackage{subfigure}

\newcommand{\q}{\mathbf q}

\newtheorem{lemma}{Lemma}

\begin{document}
\title{Energy Trade-off in Ground-to-UAV Communication via Trajectory Design}
\author{\IEEEauthorblockN{Dingcheng~Yang,~\IEEEmembership{Member,~IEEE,} Qingqing Wu,~\IEEEmembership{Member,~IEEE,} Yong~Zeng,~\IEEEmembership{Member,~IEEE,} and~Rui~Zhang,~\IEEEmembership{Fellow,~IEEE}}
\thanks{D. Yang is with the Information Engineering School, Nanchang University, Nanchang 330031, China. (e-mail: yangdingcheng@ncu.edu.cn).}
\thanks{Q. Wu, Y. Zeng, and R. Zhang are with the Department of Electrical and Computer Engineering, National University of Singapore, Singapore 117583 (e-mail:\{elewuqq, elezeng, elezhang\}@nus.edu.sg).}
\vspace{-4ex}
}

\maketitle

\begin{abstract}
Unmanned aerial vehicles (UAVs) have a great potential for improving the performance of wireless communication systems due to their wide coverage and high mobility. In this paper, we study a UAV-enabled data collection system, where a UAV is dispatched to collect a given amount of data from a ground terminals (GT) at fixed location. Intuitively, if the UAV flies closer to the GT, the uplink transmission energy of the GT required to send the target data can be more reduced. However, such UAV movement may consume more propulsion energy of the UAV, which needs to be properly controlled to save its limited on-board energy. As a result, the transmission energy reduction of the GT is generally at the cost of higher propulsion energy consumption of the UAV, which leads to a new fundamental energy trade-off in Ground-to-UAV (G2U) wireless communication. To characterize this trade-off, we consider two practical UAV trajectories, namely circular flight and straight flight. In each case, we first derive the energy consumption expressions of the UAV and GT, and then find the optimal GT transmit power and UAV trajectory that achieve different Pareto optimal trade-off between them. Numerical results are provided to corroborate our study.
\end{abstract}

\begin{IEEEkeywords}
UAV communication, energy-efficient communication, energy trade-off, trajectory design.
\end{IEEEkeywords}


\section{Introduction}
Unmanned aerial vehicles (UAVs) equipped with communication transceivers have found increasingly more applications in wireless communication, such as for information broadcasting, relaying, data collection, etc \cite{zeng2016wireless}. This is mainly attributed to the flexible deployment and high mobility of UAVs, as well as their line-of-sight (LoS) communication links with the ground terminals (GTs) at moderate altitude. There are mainly two lines of research in the existing literature on UAV-to-Ground (U2G)/Ground-to-UAV (G2U) communications, depending on whether the UAV's mobility is fully exploited or not. One line of works mainly focus on optimizing the placement/deployment of static or quasi-static UAVs, to achieve the maximum communication coverage of GTs \cite{al2014optimal,Yaliniz20163DMUAV,mozaffari2016efficient,Azari2016optimalUAV,lyu2016placement}. The other research thrust aims to fully exploit the high mobility of UAVs via their trajectory design and optimization, which brings a new degree of freedom in optimizing the performance of wireless communication systems. To this end, joint communication and UAV trajectory optimization has been studied for various wireless systems, such as mobile relaying \cite{zeng2016throughput}, multiple access channel (MAC) and broadcast channel (BC) \cite{wu2017joint}, \cite{wu2017interference}.

On the other hand, energy saving has been recognized as an important metric in designing future wireless communication systems \cite{wu2016overview}. For instance, prior works \cite{Zhan2017UAVletter,mozaffari2016optimal,Alzenad2017placement,li2016energy} have studied the energy minimization of the UAVs and/or GTs in various U2G/G2U communication systems. However, these works only focus on minimizing the communication energy consumption as in the conventional terrestrial wireless communication \cite{wu2016overview}. For UAVs in practice, their communication energy consumption is usually much lower compared to propulsion energy consumption, which is required to maintain the UAVs aloft and enable their mobility. Due to the limited on-board energy of UAVs, their propulsion energy consumption becomes the dominant factor that needs to be taken into account for achieving energy-efficient UAV communications \cite{zeng2016energy}. To this end, the authors in \cite{zeng2016energy} developed a mathematical model for the propulsion energy consumption of fixed-wing UAVs, based on which energy-efficient UAV trajectories were designed in various U2G/G2U communication systems \cite{zeng2016energy,SeongahJeong2017UAVCloud}.

\begin{figure}
\centering
\subfigure[Circular flight]{\includegraphics[width=0.35\textwidth,height=1.3in]{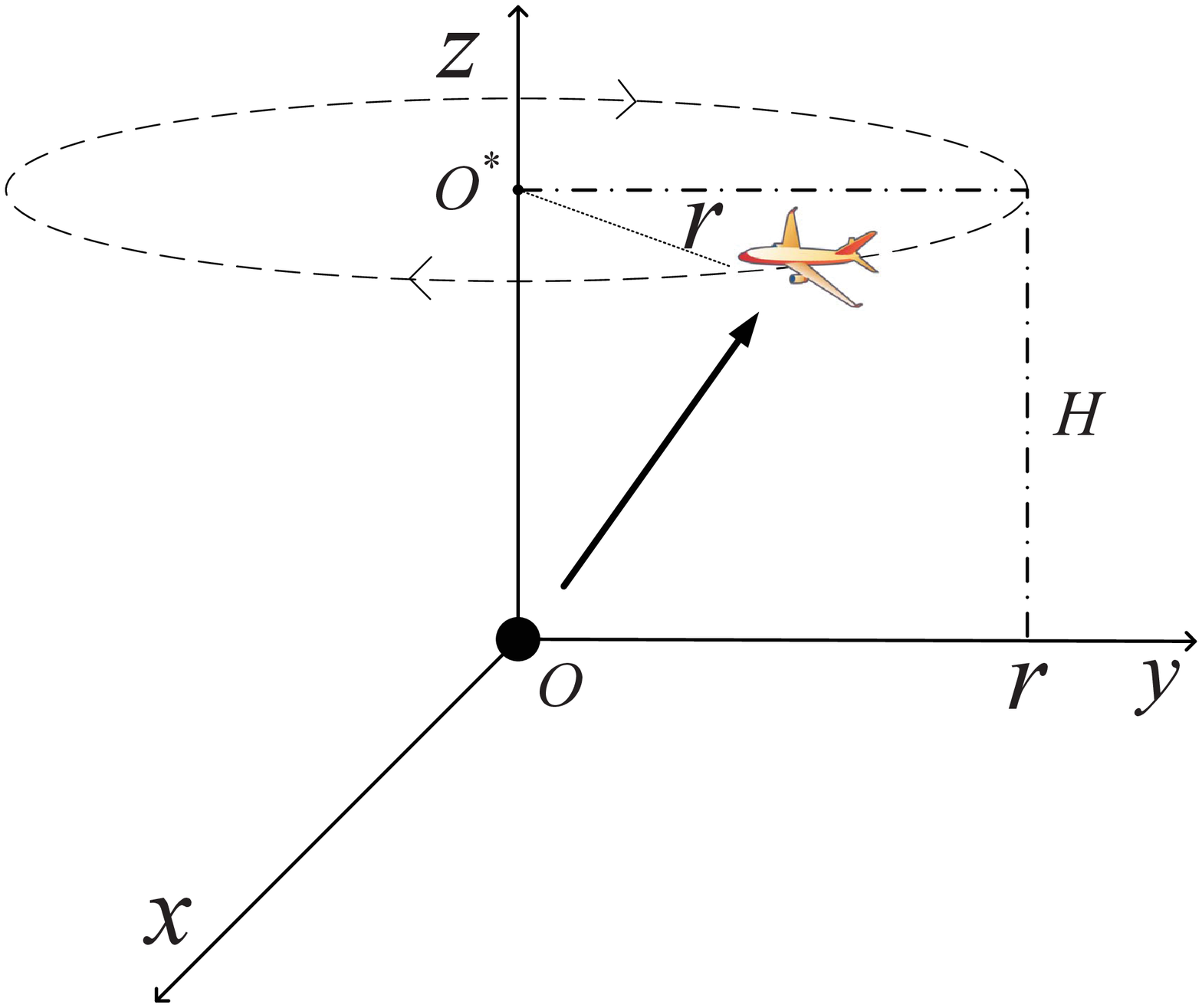}}~~~
\subfigure[Straight flight]{\includegraphics[width=0.35\textwidth,height=1.2in]{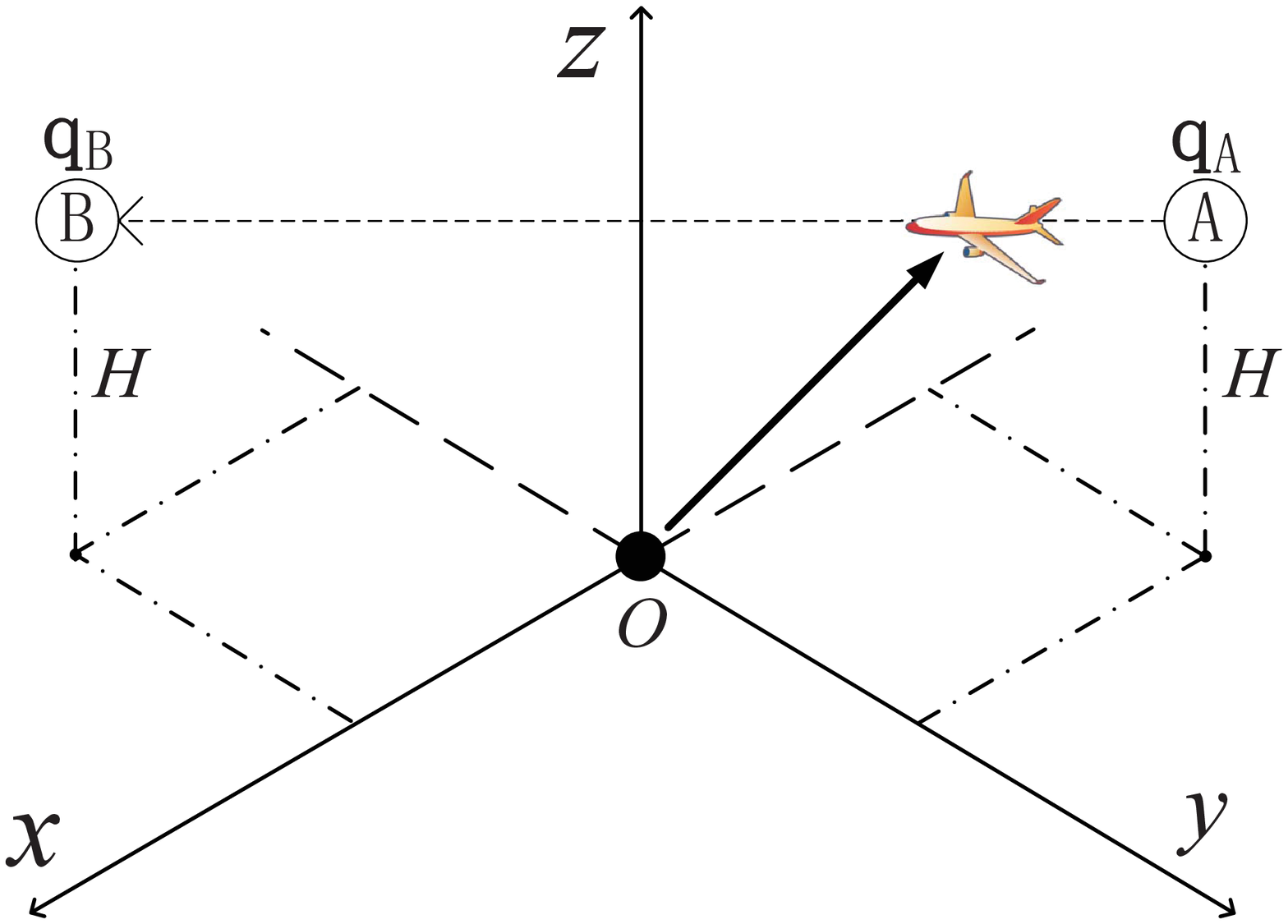}}
\caption{A ground-to-UAV wireless communication system with circular or straight flight UAV trajectory.\vspace{-3ex}} \label{model}
\end{figure}

In this paper, we study a G2U wireless communication system, where a UAV is dispatched as a mobile data collector to gather a given amount of data from a fixed GT at known location. Intuitively, the GT will consume less uplink transmission energy to send the data if the UAV can fly closer to it to establish better G2U channels \cite{Zhan2017UAVletter}. However, such movement usually requires more propulsion energy consumption of the UAV. As a result, the transmission energy reduction of the GT is generally at the cost of higher propulsion energy consumption of the UAV, which leads to a new fundamental energy trade-off in G2U wireless communication. We aim to characterize this new trade-off by taking into account the communication (including both the transmission and circuit) power consumption of the GT as well as the propulsion power consumption of the UAV. To gain the most essential insights, we focus our study on a point-to-point G2U communication system by considering two practical UAV trajectories, namely circular flight and straight flight, as shown in Fig. \ref{model}. The problem formulations and solutions in this paper can be extended to more general setups with more UAVs/GTs and arbitrary UAV trajectories, in light of the recent results on joint communication and UAV trajectory design in the literature (see, e.g., \cite{zeng2016throughput}, \cite{wu2017joint}).  For both types of trajectories, we obtain the optimal GT transmit power and UAV trajectory that achieve different Pareto optimal energy trade-off between GT and UAV. Numerical results are provided to corroborate our study.

\section{System Model}
We consider a G2U wireless communication system as shown in Fig. \ref{model}. A UAV is dispatched as a mobile data collector to gather the data of total size $Q$ bits from a GT at fixed location (e.g., a sensor node in wireless sensor network). Without loss of generality, we assume that the horizontal coordinate of the GT is ${\mathbf{w}}=[0,0]^T$. Furthermore, we assume that the UAV flies at a constant altitude $H$, which could correspond to the minimum altitude due to safety consideration. Further denote by $T$ the required mission completion time and $\mathbf {q}(t)\in \mathbb{R}^{2\times 1}$ the UAV trajectory projected onto the ground, $0\leq t\leq T$. Then the time-varying distance between the GT and UAV is given by $d(t)=\sqrt{H^2+\|\mathbf q(t)\|^2}$. We assume that the channel between the GT and the UAV is dominated by LoS link. 
 Thus, the channel power gain from the GT to the UAV at time $t$ can be modelled as
  \begin{align}
h(t)=\beta_0d^{-2}(t)=\frac{\beta_0}{H^2+||\q(t)||^2},
\end{align}
where $\beta_0$  denotes the channel gain at the reference distance $d_0=1$ meter.

We assume that throughout the time horizon $T$, the GT keeps uploading a file to the UAV with a constant transmit power $p_1$\footnote{For simplicity, in this paper we do not consider adaptive power control of the GT based on the instantaneous channel $h(t)$, which can be employed to further improve the energy efficiency of the GT.}, where $p_1\leq \bar{P}_1$ with $\bar{P}_1$ denoting the maximum allowable transmission power at the GT. As a result, the total amount of information bits that can be uploaded to the UAV is a function of the time horizon $T$, GT transmit power $p_1$, and the UAV trajectory $\mathbf q(t)$, which can be expressed as
\begin{equation}\label{eq44}
\begin{aligned}
\bar Q(T,p_1,\q(t))=&B \int_0^T \log_2\left(1+\frac{p_1h(t)}{\sigma^2}\right)dt\\
=& B \int_0^T \log_2\left(1+\frac{p_1\gamma_0}{H^2+||\q(t)||^2}\right)dt,
\end{aligned}
\end{equation}
where $B$ is the system bandwidth in hertz (Hz), $\sigma^2$ denotes the noise power at the UAV receiver, and $\gamma_0\triangleq\frac{\beta_0}{\sigma^2}$.

Furthermore, the total energy consumption of the GT to complete the file upload can be expressed as
\begin{align}\label{E1}
\bar E_{1}(T,p_1)=T(p_1+P_{\text{c}}),
\end{align}
where $P_{\text{c}}$ denotes the constant circuit power of the GT \cite{wu2016overview}.

On the other hand, the energy consumption of the UAV mainly consists of two parts. The first part is the communication-related energy consumption for circuitry and signal processing. The other part is the propulsion energy consumption, which is required for the UAV to remain aloft and move freely. In practice, the communication-related energy is much smaller than the propulsion energy, and is thus ignored for the UAV in this paper. As derived in \cite{zeng2016energy}, the propulsion energy consumption of fixed-wing UAVs is a function of its trajectory $\q(t)$, which can be expressed in a generic form as
\begin{align}\label{E3}
\bar E_{2}(T,\q(t))=   \int_0^T p_{2}(\q(t))dt,
\end{align}
where $p_2(\q(t))$ represents the UAV's instantaneous propulsion power consumption that depends on the its flying status at time instant $t$. Denote by  $\bar P_2$ the maximum propulsion power that can be provided by the UAV. We thus have $p_{2}(\q(t))\leq \bar P_2$, $\forall t$.
 It is observed from \eqref{eq44}--\eqref{E3} that the aggregated achievable communication throughput $\bar Q$, the GT energy consumption $\bar E_1$, and the UAV energy consumption $\bar E_2$ are closely coupled via the three design parameters $T$, $p_1$, and $\q(t)$. In the following, we investigate the required $E_1$ and $E_2$ for given  $\bar Q$.

\section{GT-UAV Energy Trade-off}
For any given target file size $Q$ to be uploaded by the GT, we define the GT-UAV energy consumption region $\mathcal C$ as the union of all the feasible energy pairs $(E_1, E_2)$ that are sufficient to complete the file upload. Mathematically, we have
\begin{align}\label{eq9}
\mathcal{C} \triangleq  &  \mathop{\bigcup}\limits_{{T, p_1, \q(t)}}  \Bigg\{ (E_1, E_2):   \nonumber \\
&E_1= T(p_{1}+P_{\rm{c}}), E_{\rm 2} =  \int_0^T p_{\rm 2}(\q(t)) dt,  \nonumber\\
& \bar Q(T,p_1,\q(t))\geq Q, p_{\rm 1}\leq \bar P_1,   p_{\rm 2}(\q(t)) \leq \bar P_{2}, \forall t \Bigg \}.
 \end{align}
Of particular interest is the Pareto boundary of $\mathcal C$, which is defined as the set of all energy pairs $(E_1, E_2)$ at which it is impossible to decrease one of them without increasing the other. The corresponding design of $(T, p_1, \q(t))$ is referred to as a Pareto optimal solution. \begin{lemma}
At the Pareto optimal solution, we always have
\begin{align}\label{eq:StrictEq}
\bar Q(T,p_1,\q(t))=Q.
\end{align}
\end{lemma}
\begin{IEEEproof}
Lemma 1 can be shown by contradiction. Suppose that at the Pareto optimal solution we have $\bar Q(T, p_1,\q(t))>Q$, then we can reduce the operation time $T$ so that the energy consumption  of both the GT and UAV are reduced. Since this violates the Pareto optimality, the lemma is proved.
\end{IEEEproof}

In the following two subsections, we study the Pareto boundary of $\mathcal C$ by considering two types of UAV trajectories that are of high practical interests. The first one is the circular flight where the UAV flies around the GT following a circular trajectory. The second one is steady straight flight, where the UAV flies from a given initial location $\q_A\in \mathbb{R}^{2 \times 1}$ to a final location $\q_B\in \mathbb{R}^{2 \times 1}$ along the line connecting them with a constant speed.

\subsection{Circular Flight}
For fixed-wing UAVs, circular flight is a practical trajectory that offers a flexible trade-off between UAV's energy consumption and the communication throughput via designing the circle radius $r$ and flying speed $V$ \cite{zeng2016energy}. In this section, we characterize the new GT-UAV energy trade-off with circular UAV trajectory, as shown in Fig. \ref{model}(a). First, it is noted that with a circular trajectory of radius $r$ and the GT at the center of its projection on the ground, we have $\|\q(t)\|=r$, $\forall t$. As a result, the communication throughput \eqref{eq44} reduces to
\begin{align}\label{eq3}
  \bar Q(T, p_1, r)= B T\log_2\left(1+\frac{p_1\gamma_0}{H^2+r^2}\right).
\end{align}
Furthermore, it follows from \cite{zeng2016energy} that for fixed-wing UAV with circular trajectory of  radius $r$ and constant speed $V$, the required instantaneous propulsion power can be expressed as
\begin{align}
p_2(\q(t))=p_2(r,V)=\left(c_1+\frac{c_2}{g^2r^2}\right)V^3+\frac{c_2}{V}, \label{eq:p2rv}
\end{align}
where $c_1$ and $c_2$ are the parameters depending on the aircraft's weight, wing area, air density, etc. It is observed from \eqref{E1} and \eqref{eq3} that for circular UAV flight, the UAV speed $V$ neither affects the GT energy consumption nor the communication throughput. Thus, it should be chosen to minimize the UAV's energy consumption, or equivalently the instantaneous UAV propulsion power $p_2(r,V)$ in \eqref{eq:p2rv}. For any given circular radius $r$,  the optimal UAV speed for minimizing \eqref{eq:p2rv} can be obtained as \cite{zeng2016energy}
\begin{equation}
V^\star(r) = \left(\frac{c_2}{3\left(c_1+c_2/\left(g^2r^2\right)\right)}\right)^{1/4},
\end{equation}
and the corresponding minimum UAV propulsion power is
\begin{equation}\label{eq:p2r}
p^\star_{2}(r)= \left(3^{-3/4}+3^{1/4}\right)c^{3/4}_2\left(c_1+\frac{c_2}{g^2r^2}\right)^{1/4}.
\end{equation}
Therefore, with circular trajectory, the UAV's energy consumption reduces to a function of the mission completion time $T$ and the circle radius $r$, i.e.,
\begin{align}
\bar E_{2}(T,r) = Tp^\star_{2}(r).\label{eq:E2Circ}
\end{align}
Note that to ensure the maximum UAV propulsion power constraint  $p_2^\star(r)\leq \bar P_2$ is satisfied, it follows from \eqref{eq:p2r} that we must have $\bar P_2 > (3^{-3/4}+3^{1/4})c_1^{1/4}c_2^{3/4}$ (by taking $r\rightarrow \infty$ in \cite{wu2016overview}), and the UAV's trajectory radius cannot be smaller than a certain threshold, i.e.,
\begin{align}
 r\geq r_{\min} \triangleq \frac{1}{g}\sqrt{\frac{c_2}{ \frac{\bar{P}^4_2}{(3^{-3/4}+3^{1/3})^4c_2^{3}} - c_1   }}. \label{eq:rMin}
\end{align}
To characterize the Pareto boundary $(E_1, E_2)$ of $\mathcal C$ under circular trajectory, we obtain the following set of equations based on \eqref{E1}, \eqref{eq:StrictEq}, \eqref{eq3}, and \eqref{eq:E2Circ}:
\begin{numcases}{}
E_1=T(p_1+P_{\text{c}}),\label{eq:eq1}\\%
E_2=Tp_2^\star(r),\label{eq:eq2}\\ %
B T\log_2\left(1+\frac{p_1\gamma_0}{H^2+r^2}\right)=Q,\label{eq:eq3}\\
r\geq r_{\min},\label{eq:eq4}\\
p_1\leq \bar{P}_1.  \label{eq:eq5}
\end{numcases}
It can be observed from \eqref{eq:eq2} that for fixed $T$, the decrease of $E_2$ implies the increase of $r$. As a result, it follows from \eqref{eq:eq3} that, $p_1$ needs to be increased and hence $E_1$ in \eqref{eq:eq1} will increase. This clearly suggests that there exists a  GT-UAV energy trade-off for any given operation time $T$. Before characterizing the complete Pareto boundary of $\mathcal C$ based on \eqref{eq:eq1}--\eqref{eq:eq5}, we first obtain its two extreme points corresponding to the minimum energy consumption of either the UAV or the GT, which are denoted by $(E_{1, {\min}}, E_{2, {\max}})$ and $(E_{1, {\max}}, E_{2, {\min}})$, respectively.   By eliminating $T$ based on \eqref{eq:eq1} and \eqref{eq:eq3},  $E_{1, {\min}} $ can be obtained by solving the following optimization problem
\begin{align}\label{eq:eq444}
E_{1, {\min}} = & \min \limits_{p_1, r}  \frac{Q(p_1+P_{\text{c}})}{B\log_2\left(1+\frac{p_1\gamma_0}{H^2+r^2}\right)}, \nonumber\\
&\text{s.t.} ~~ p_1\leq \bar P_1, r\geq r_{\min}.
\end{align}
It follows that at the optimal solution to \eqref{eq:eq444}, we should have $r=r_{\min}$. In other words, from the perspective of minimizing the GT's energy consumption, the UAV should hover around the GT with the minimum possible circle radius, as expected. As a result, \eqref{eq:eq444} reduces to the conventional power optimization problem for energy-efficient point-to-point communication \cite{miao2013energy1}, which is a quasiconvex problem and hence can be efficiently solved with bisection search over $p_1$. In the special case that $P_{\text{c}}=0$, the optimal power allocation for GT energy minimization is $p_1\rightarrow 0$, as shown in \cite{miao2013energy1}. In this case, it takes an infinitely long time to complete the file uploading.
With $r$ and $p_1$ obtained from \eqref{eq:eq444}, the corresponding energy consumption of the UAV, denoted as $E_{2, \max}$, can be accordingly computed from \eqref{eq:eq2} and \eqref{eq:eq3}.
Similarly, to obtain the other extreme point where the UAV achieves the minimum energy consumption, we have the following optimization problem based on \eqref{eq:eq2} and \eqref{eq:eq3}
\begin{align}\label{eq:eq555}
E_{2, {\min}} =  &\min \limits_{p_1, r}  \frac{Q p^\star_2(r)}{B\log_2(1+\frac{p_1\gamma_0}{H^2+r^2})}, \nonumber \\
&\text{s.t.} ~~ p_1\leq \bar P_1, r\geq r_{\min}.
\end{align}
 At the optimal solution to \eqref{eq:eq555}, we should have $p_1=\bar P_1$. As a result, \eqref{eq:eq555} reduces to a univariate optimization problem with respect to $r$,  which can be solved via one-dimensional search. With $r$ and $p_1$ from \eqref{eq:eq555}, $E_{1, \max}$ can be accordingly computed from \eqref{eq:eq1} and \eqref{eq:eq3}.

Next, we characterize the complete Pareto boundary of $\mathcal C$ based on \eqref{eq:eq1}--\eqref{eq:eq5}. To this end, by first solving $r$ and $p_1$ based on \eqref{eq:eq2} and \eqref{eq:eq3}, we have
 \begin{align}\label{eq:rE2T}
 r=r(E_2,T) \triangleq \frac{1}{g}\sqrt{\frac{c_2}{ \frac{E^4_2}{T^4(3^{-3/4}+3^{1/3})^4c_2^{3}} - c_1}} \geq r_{\min},
 \end{align}

\begin{align}
p_1=p_1(E_2, T)\triangleq \frac{H^2+r^2(E_2,T)}{\gamma_0}\left( 2^{\frac{{Q}}{BT}}-1\right) \leq \bar{P}_1.\label{eq:p1E2r}
\end{align}
By substituting the above $r$ and $p_1$ into \eqref{eq:eq1}, we have
\begin{align}\label{eq:fE2r}
&E_{1}= f(E_2, T) \triangleq  \nonumber\\
&T\left[\left(H^2+\frac{1}{g^2}\frac{c_2}{ \frac{E^4_2}{T^4(3^{-3/4}+3^{1/3})^4c_2^{3}} - c_1}\right)\frac{2^{\frac{Q}{BT}}-1}{\gamma_0}+P_{\rm c}\right].
\end{align}
In other words, for circular UAV flight, any Pareto optimal energy pair $(E_1, E_2)$ must satisfy the closed-form expression given by \eqref{eq:fE2r}, which is parameterized by the UAV operation time $T$. For any given $T$, the function $E_1=f(E_2, T)$ is a monotonically decreasing function with respect to $E_2$, which clearly shows the trade-off between $E_1$ and $E_2$. Furthermore, to satisfy the UAV's propulsion power constraint, the operation time $T$ must be in the interval
\begin{align}\label{eq:Tregion}
  T \in \left[\frac{E_2}{\bar{P}_2},\frac{E_2}{(3^{-3/4}+3^{1/4})c_1^{1/4}c_2^{3/4}} \right).
\end{align}

To characterize the complete Pareto boundary of $\mathcal C$, for any given $E_2\in [E_{2,\min}, E_{2,\max}]$, the energy consumption of the GT $E_1$ is minimized based on \eqref{eq:fE2r} by optimizing the operation time $T$, i.e.,
  \begin{equation}\label{probm7}
 \begin{aligned}
\min  \limits_{T}&  ~~ E_{1}= f(E_2, T) \\
\text{s.t.} &~~ \   \eqref{eq:rE2T}, \eqref{eq:p1E2r},\eqref{eq:Tregion}. 
 \end{aligned}
 \end{equation}
Problem \eqref{probm7} is a univariate optimization problem, which can be efficiently solved via one-dimensional search.

 \subsection{Straight Flight}
In this subsection, we consider another practical trajectory, where the UAV flies from a given initial location $\q_A$ to a final location $\q_B$ with constant speed $V$, as shown in Fig. \ref{model}(b). In this case, the operation time of interest is $T=\frac{D}{V}$, where $D=\|\q_B-\q_A\|$ is the distance between $\q_A$ and $\q_B$. Furthermore, the UAV's trajectory $\q(t)$ can be expressed as
 \begin{align}
 \q(t) =\q_A  + t V\vec {\mathbf{d}}, \ 0 \leq t \leq \frac{D}{V},
 \end{align}
 where $\vec{\mathbf{d}}=\frac{\q_B-\q_A}{D}$ is the unit-norm vector representing the flying direction from $\q_A$ to $\q_B$. As a result, the time-varying distance between the GT and the UAV can be expressed as
 \begin{align}
 d(t)& =\sqrt{ \| \q(t) \|^2+H^2}  
 =\sqrt{(Vt+c_3)^2 +  \overline{H}^2},
 \end{align}
 where $c_3 \triangleq \q_A^T \vec{\mathbf{d}}$ and $ \overline{H}^2 \triangleq \|\q_A\|^2-c_3^2+ H^2$. It then follows from (\ref{eq44}) that the aggregated throughput with straight UAV flight reduces to a bivariate  function in terms of $V$ and $p_1$, which can be expressed as
\begin{align}\label{eq14}
\bar Q(p_1, V) 
=&B\int_0^{D/V} \log_2\left(1+\frac{p_1\gamma_0}{\overline{H}^2+ (Vt+c_3)^2}  \right)dt,\nonumber\\
=&\frac{BG(p_1)}{V\ln2}, 
\end{align}
where $G(p_1) \triangleq F(D+c_3)-F(c_3)$ with
\begin{equation}
\small
\begin{aligned}\label{eq:integralFormula}
F(z)&\triangleq \int \ln\left(1+\frac{p_1\gamma_0}{\overline{H}^2+z^2}\right)dz \\
&=z \ln \left( 1+\frac{p_1\gamma_0}{\overline{H}^2+z^2}\right)-2\overline{H}\tan^{-1}\left(\frac{z}{\overline{H}}\right)\\
&+2\sqrt{\overline{H}^2+p_1\gamma_0}\tan^{-1}\left(\frac{z}{\sqrt{\overline{H}^2+p_1\gamma_0}}\right).
\end{aligned}
\end{equation}
It can be verified that $G(p_1)$ is a strict monotonically increasing function, and hence its inverse function exists,  which is denoted as $G^{-1}(\cdot)$.
Furthermore, it follows from \cite{zeng2016energy} that for a fixed-wing UAV with steady straight flight of constant speed $V$, the required instantaneous propulsion power can be expressed as $p_2(\q(t))=p_2(V)=c_1V^3+\frac{c_2}{V}$, where  $c_1$ and $c_2$ are the same as those in (\ref{eq:p2rv}). As a result, the UAV's energy consumption reduces to a univariate function in terms of the speed $V$, i.e.,
\begin{align}\label{E5}
\bar E_{2}(V) = \frac{D}{V}p_{2}(V)= D\left( c_1V^2+\frac{c_2}{V^2}\right).
\end{align}

To characterize the Pareto boundary of $\mathcal C$ with straight trajectory, we obtain the following set of equations based on  \eqref{E1}, \eqref{eq:StrictEq}, \eqref{eq14}, and \eqref{E5}:
\begin{numcases}{}
E_1=\frac{D}{V}(p_1+P_{\text{c}}), \label{eq:eq11}\\
E_2= D\left( c_1V^2+\frac{c_2}{V^2}\right),\label{eq:eq22} \\
\frac{BG(p_1)}{V\ln2}=Q, \label{eq:eq33}\\
p_1 \leq \bar P_1,\label{eq:eq44}\\
p_2(V) \leq \bar P_2.\label{eq:eq55}
\end{numcases}
From \eqref{eq:eq22}, it can be inferred that there exists one unique speed $V$ for minimizing $E_2$. Meanwhile, it follows  from  \eqref{eq:eq33} that for any given $Q$, $p_1$ increases with $V$. As such, from \eqref{eq:eq11}, there exists another speed $V$ for minimizing $E_1$, which is in general different from that for minimizing $E_2$. Thus, $V$ must be properly chosen to balance the energy consumptions of the UAV and GT.
Similar to the circular trajectory case,  we first derive the two extreme points $(E_{1, {\min}}, E_{2,{\max}})$ and $(E_{1, {\max} }, E_{2, {\min}})$ in the UAV-GT energy trade-off. By eliminating $V$ based on \eqref{eq:eq11} and \eqref{eq:eq33}, $E_{1, {\min}}$ can be obtained by solving the  following optimization problem
\begin{align}\label{eq:eqE1min}
E_{1, {\min}} =  &\min \limits_{p_1} \frac{DQ\ln2}{BG(p_1)}(p_1+P_{\text{c}}), \nonumber\\
&~\text{s.t.} ~~ p_1\leq \bar P_1, p_2\left(\frac{BG(p_1)}{Q\ln2}\right)\leq \bar P_2.
\end{align}
Then, $E_{1, {\min}}$ can be obtained by one-dimensional search over $p_1$. The corresponding $E_{2, \max}$ can be obtained from \eqref{eq:eq22} and \eqref{eq:eq33} with  $p_1$ given in \eqref{eq:eq44}. Similarly, by eliminating $V$ based on  \eqref{eq:eq22} and \eqref{eq:eq33}, $E_{2,\min}$ can be obtained by solving the following optimization problem
\begin{align}\label{eq:eq56}
E_{2, {\min}} = & \min \limits_{p_1} D\left(    c_1\left(\frac{BG(p_1)}{Q\ln2}\right)^2+c_2\left(\frac{Q\ln2}{BG(p_1) }\right)^2      \right) \nonumber\\
&~\text{s.t.} ~~ p_1\leq \bar P_1, p_2\left(\frac{BG(p_1)}{Q\ln2}\right)\leq \bar P_2.
\end{align}
Then, $E_{2, {\min}}$ can be obtained by one-dimensional search over $p_1$. The corresponding $E_{1,\max}$ can be obtained from \eqref{eq:eq11} and \eqref{eq:eq33} with  $p_1$ in \eqref{eq:eq56}.

Next, we characterize the complete Pareto boundary of $\mathcal C$ based on \eqref{eq:eq11}--\eqref{eq:eq55}. To this end, for any given $E_2$, we aim to minimize $E_1$ while ensuring that all conditions in  \eqref{eq:eq11}--\eqref{eq:eq55} are satisfied.
By solving $p_1$ from \eqref{eq:eq33}, we obtain $p_1 = G^{-1}\left(\frac{QV\ln2}{B}\right)$. In addition, it follows from \eqref{eq:eq22} that for any given $E_2 \geq 2 D\sqrt{c_1c_2}$, there exist two different values of the UAV speed $V$ that achieve the same energy consumption $E_2$, which are given by
\begin{align}
&V_1= g_1(E_2) \triangleq \sqrt{\frac{E_2 +\sqrt{E_2^2-4c_1c_2D^2}}{2Dc_1}}\label{eq:eq7777},\\
&V_2= g_2(E_2) \triangleq  \sqrt{\frac{E_2 - \sqrt{E^2_2-4c_1c_2D^2}}{2Dc_1}}.\label{eq:eq8888}
\end{align}
Note that to ensure that \eqref{eq:eq44} and \eqref{eq:eq55} are satisfied, we have the following equivalent conditions with respect to $E_2$:
\begin{align}
&G^{-1}\left(\frac{Qg_k(E_2)\ln2}{B}\right) \leq \bar P_1, k =1,2, \label{eq:eq91}\\
&D\left( c_1g^2_k(E_2)+\frac{c_2}{g^2_k(E_2)}\right) \leq  \bar P_2,  k =1,2.\label{eq:eq92}
\end{align}
By substituting $p_1$ and $V=g_k(E_2), k =1,2$, into \eqref{eq:eq11}, we obtain two possible values for $E_1$ for any given $E_2$, which are given in closed-form as
\begin{align}\label{eq:fE2r2}
E_{1,k} = \frac{D}{g_k(E_2)   }\left[G^{-1}\left(\frac{Qg_k(E_2)\ln2}{B}\right)+P_{\rm c}\right], k=1, 2.
\end{align}
As a result, for any given $E_2\in [E_{2,\min}, E_{2,\max}]$ that satisfies both \eqref{eq:eq91} and \eqref{eq:eq92},
the corresponding Pareto optimal point is obtained by simply comparing the two GT energy consumption values in \eqref{eq:fE2r2}, i.e., $E_1=\min \{E_{1,1}, E_{1,2}\}$.

\section{Numerical Results}
\begin{figure}
\centering
\includegraphics[width=0.5\textwidth]{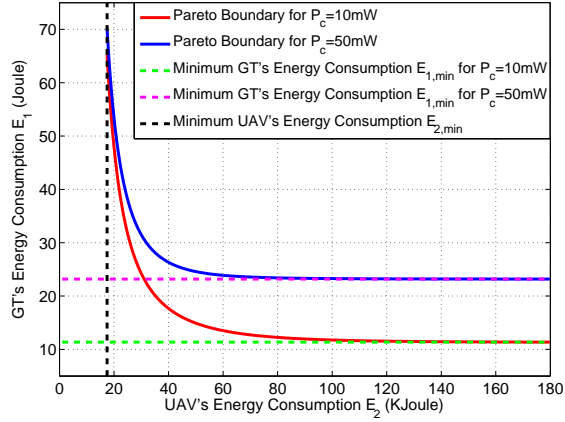}
\caption{GT-UAV energy trade-off with circular trajectory and $P_{\text{c}}=$ 50mW or 10mW .}\label{circular}
\end{figure}
This section provides numerical results to verify the GT-UAV energy trade-off developed in this paper. Unless otherwise stated, the parameters are set as follows: $H=100$ m, $B=1$ MHz, $\sigma^2=-110$ dBm, $\gamma_0=-50$ dB, $c_1=9.26\times 10^{-4},c_2=2250$, $\bar P_2=1500$ W, and $\bar P_1=0.5$ W.

We first consider the circular flight, where the amount of data to be collected by the UAV is $\bar{Q}=600$ Mbits(Mb). The energy trade-off curves between the GT and UAV are shown in Fig. \ref{circular} for two different circuit power levels $P_{\text{c}}=50$mW and $10$mW. It is observed that for both cases, as the UAV's energy consumption $E_2$ increases, the GT's energy consumption $E_1$ decreases correspondingly, which demonstrates the fundamental trade-off for the GT-UAV energy consumptions, and such a trade-off relationship is more substantial for smaller $P_{\text{c}}$.  For example, for $P_{\text{c}}=10$mW, by increasing the UAV's energy from 18 KJ to 40 KJ, the GT energy consumption would be significantly reduced by about 3-4 times. This is beneficial for energy-limited GTs.
\begin{figure}[H]
\centering
\subfigure[$\bar{Q}=30$Mb]{\includegraphics[width=0.5\textwidth]{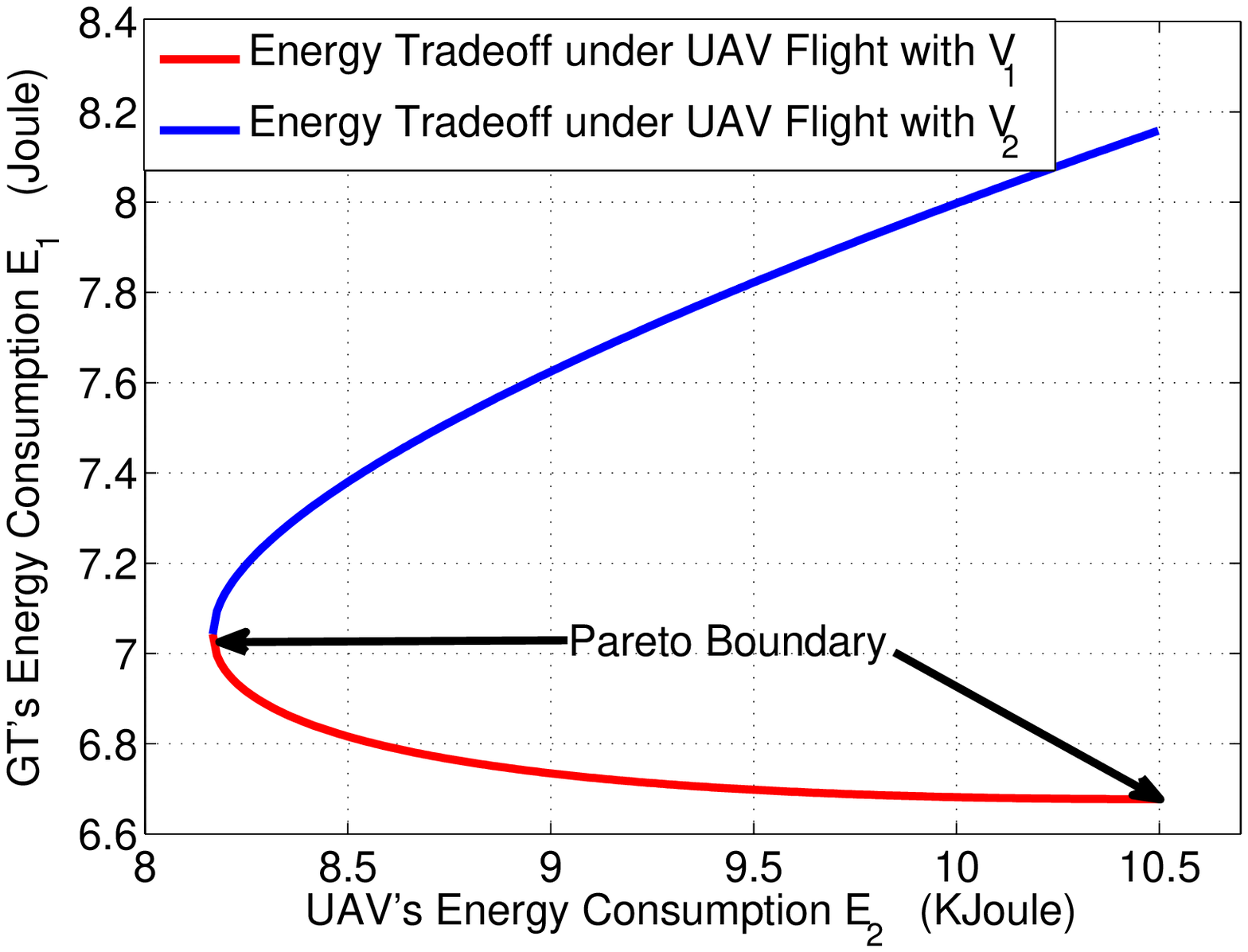}}~~~
\subfigure[$\bar{Q}=100$Mb]{\includegraphics[width=0.5\textwidth]{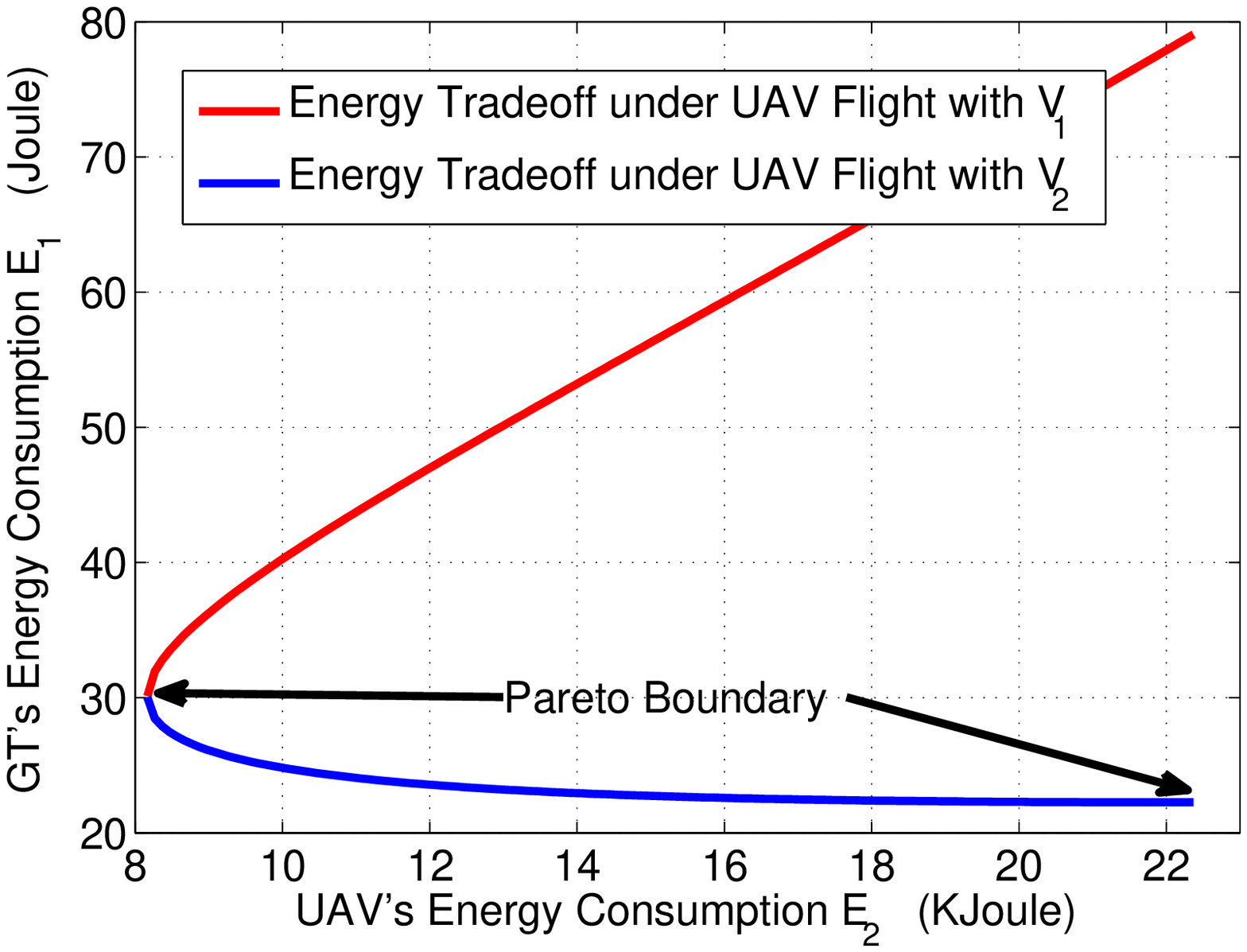}}
\caption{GT-UAV energy trade-off with straight trajectory and $\bar{Q}=$ 30Mb or 100Mb.} \label{straightflight}
\end{figure}
\begin{figure}[H]
\centering
\subfigure[$\bar{Q}=30$Mb]{\includegraphics[width=0.5\textwidth]{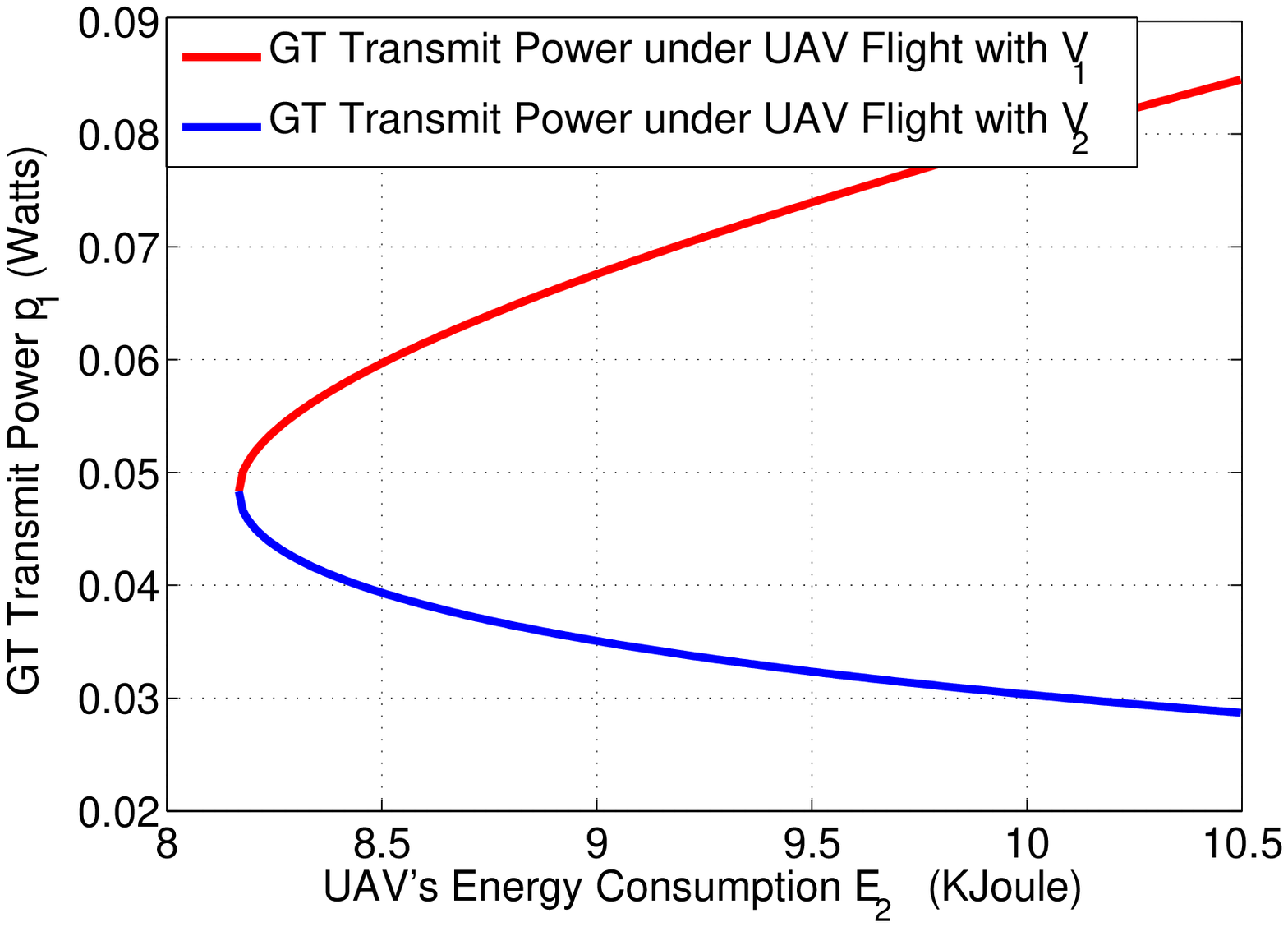}}~~~
\subfigure[$\bar{Q}=100$Mb]{\includegraphics[width=0.5\textwidth]{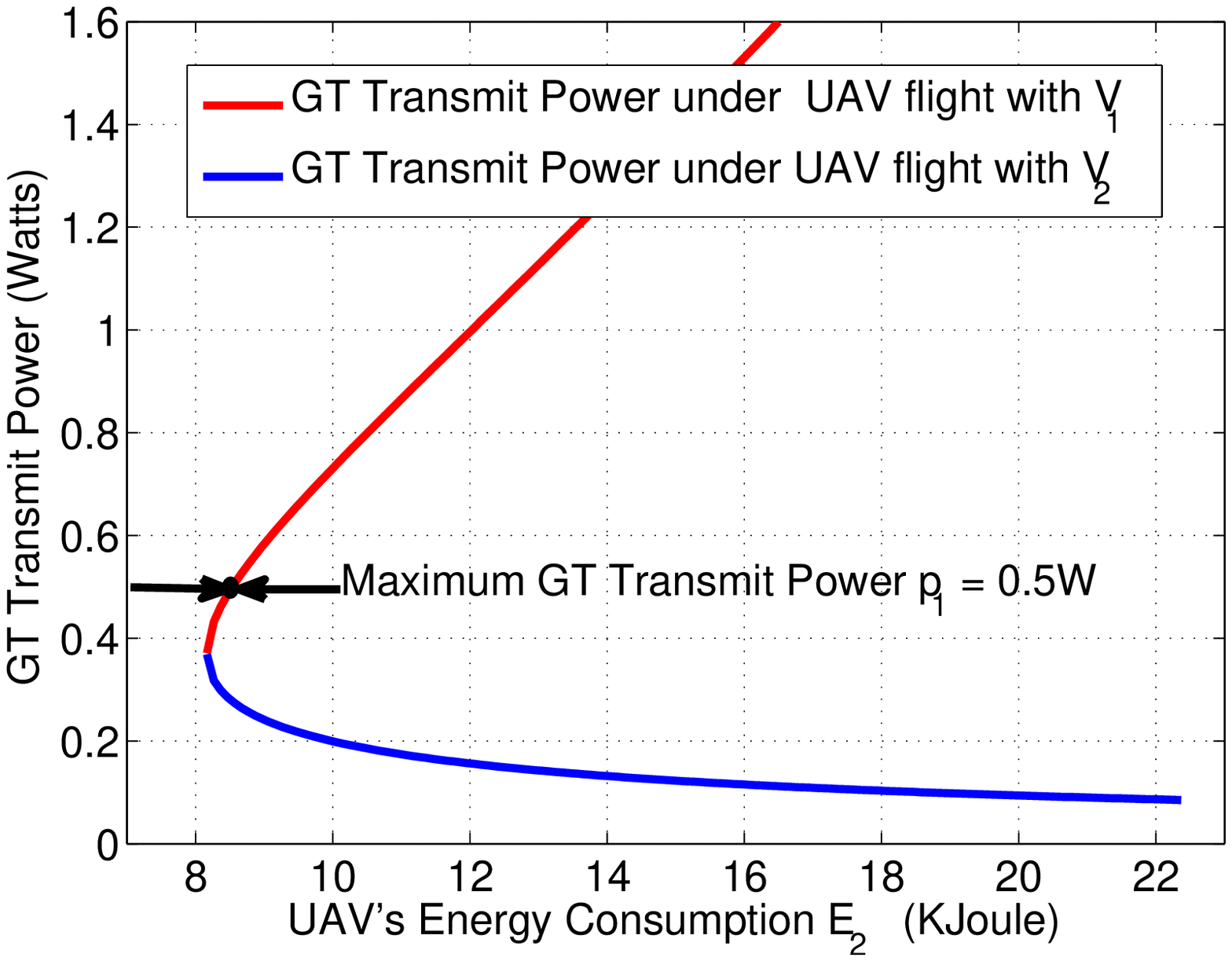}}
\caption{GT transmit power with straight trajectory and $\bar{Q}=$ 30 Mb or 100 Mb} \label{GTtransmitpowerstraightflight}
\end{figure}
Next, we consider the case of straight flight, where the UAV's initial and final locations are respectively set as $\q_A= [-1000,1000]^T$ and $\q_B=[1000,-1000]^T$ in meter, $P_{\text{c}}=50$mW and $\bar{Q}$ is set as 30 Mb or 100 Mb. The energy trade-off curves for these two different data collection requirements are shown in Fig. \ref{straightflight}. Note that in the figure, $V_1$ and $V_2$ denote the two different UAV speeds that both result in the same UAV energy consumption, as given in (\ref{eq:eq7777}) and (\ref{eq:eq8888}), respectively. It is observed that the two different UAV flight speeds can achieve the Pareto boundary under different data transmission requirements. Specifically, for $\bar{Q}=30$ Mb, it is observed that the larger speed $V_1$ gives the Pareto boundary as shown in Fig. 3(a). This is expected since for this setup, the circuit power of the GT (i.e. $P_c=50$ mW) is comparable to its transmit power, as shown in Fig. \ref{GTtransmitpowerstraightflight}(a). As a result, the larger UAV flight speed $V_1$ essentially decreases the required UAV traveling and file upload time, and hence saves the significant circuit energy consumption of the GT. On the other hand, for the case with larger $\bar{Q}=100$ Mb, the GT's transmit power is much larger than its circuit power, as shown in Fig. \ref{GTtransmitpowerstraightflight}(b). This implies that the GT's transmission energy in this case is more significant than its circuit energy consumption, and should have a higher priority to be minimized. Thus, the UAV should fly with a lower speed $V_2$ so as to have a longer transmission time, and hence lower transmit power. Thus, in this case, $V_2$ corresponds to the Pareto-optimal speed as shown in Fig. \ref{straightflight}(b).
\section{Conclusions}
In this paper, we investigate a new fundamental energy trade-off between the UAV and its served GT in a G2U wireless communication system. By deriving the propulsion energy consumption of the UAV and the communication energy consumption of the GT under two practical UAV trajectories, namely circular flight and straight flight, we characterize their Pareto optimal trade-offs and the corresponding optimal GT transmit power and UAV trajectory designs. We hope that the results of this paper shed new light on designing energy-efficient UAV communications in future wireless systems.

\bibliographystyle{IEEEtran}
\bibliography{IEEEabrv,mybib}

\end{document}